\documentclass[pra,twocolumn,nofootinbib]{revtex4-2}
\pdfoutput=1

\usepackage{subfiles}

\usepackage{graphicx}
\usepackage{hyperref}
\usepackage[section]{placeins}
\usepackage{amsmath}
\usepackage{amsthm}
\usepackage{amssymb}
\usepackage{dsfont}
\usepackage{bbold}
\usepackage{comment}
\usepackage{color}
\newtheorem{theorem}{Theorem}

\newtheorem{example}{Example}
\newcommand{\Tr}{\text{Tr}}
\newcommand\id{\leavevmode\hbox{\small1\kern-3.3pt\normalsize1}}

\newcommand{\ack}{\subsection*{\normalsize \sf \textbf{Acknowledgement}}}

\newcommand{\bit}{\begin{itemize}}
	\newcommand{\eit}{\end{itemize}\par\noindent}
\newcommand{\ben}{\begin{enumerate}}
	\newcommand{\een}{\end{enumerate}\par\noindent}
\newcommand{\beq}{\begin{equation}}
	\newcommand{\eeq}{\end{equation}\par\noindent}
\newcommand{\beqa}{\begin{eqnarray*}}
	\newcommand{\eeqa}{\end{eqnarray*}\par\noindent}
\newcommand{\beqn}{\begin{eqnarray}}
	\newcommand{\eeqn}{\end{eqnarray}\par\noindent}

\usepackage{hyperref}
\usepackage{amsmath}
\usepackage{amsthm}
\usepackage{graphicx}
\usepackage{braket}

\newtheorem{definition}{Definition}

\newtheorem*{theorem*}{Theorem}

\usepackage{braket}
\usepackage{amsmath}
\usepackage{amssymb}
\usepackage{amsmath}
\usepackage{amsmath}
\usepackage{graphicx}
\usepackage{physics}

\begin{document}

\title{Qualitative equivalence between incompatibility and Bell nonlocality}

\author{Shiv Akshar Yadavalli}
\email{sy215@duke.edu}
\affiliation{Department of Physics, Duke University, Durham, North Carolina, USA 27708}

\author{Nikola Andrejic}
\email{nikola.andrejic@pmf.edu.rs}
\affiliation{University of Ni\v s, Faculty of Sciences and Mathematics,  Vi\v segradska 33, 18000 Ni\v s, Serbia}

\author{Ravi Kunjwal}
\email{quaintum.research@gmail.com}
\affiliation{Universit\'e libre de Bruxelles, QuIC, Brussels, Belgium}
\affiliation{Aix-Marseille University, CNRS, LIS, Marseille, France}

\begin{abstract}	
Measurements in quantum theory can fail to be jointly measurable. 
Like entanglement, this incompatibility of measurements is necessary but not sufficient for violating Bell inequalities. The (in)compatibility relations among a set of measurements can be represented by a joint measurability structure,\textit{ i.e.}, a hypergraph whose vertices denote measurements and hyperedges denote all and only compatible sets of measurements. Since incompatibility is necessary for a Bell violation, the joint measurability structure on each wing of a Bell experiment must necessarily be non-trivial, \textit{i.e.,} it must admit a subset of incompatible vertices. Here we show that for any non-trivial joint measurability structure with a finite set of vertices, there exists a quantum realization with a set of measurements that enables a Bell violation,\textit{ i.e.}, given that Alice has access to this incompatible set of measurements, there exists a set of measurements for Bob and an entangled state shared between them such that they can jointly violate a Bell inequality. Hence,  a non-trivial joint measurability structure is not only necessary for a Bell violation, but also sufficient.
\end{abstract}
\maketitle

Unlike pre-quantum physical theories, measurements are at the heart of  quantum theory. It is through measurements that the Hilbert space picture of state vectors and unitary dynamics makes contact with experimental statistics (via the Born rule). These measurements are unlike any in classical physics because of the generic impossibility of implementing them simultaneously, \textit{i.e.,} they exhibit incompatibility. When combined with  shared entanglement, the incompatibility of quantum measurements powers the violation of Bell inequalities. Such Bell violations are a strong form of nonclassicality, ruling out the possibility of local hidden variable (LHV) models of physics \cite{Bell64, Bell66, CHSH, Belltest1, Belltest2, Belltest3, Spekkens16}. 

The experimental setup needed to violate a Bell inequality is called a Bell scenario \cite{BCP14}. It consists of at least two spacelike separated parties that share an entangled state and implement local measurements on their share of the state. While entanglement and incompatibility are necessary for Bell inequality violations, they are not sufficient. We know that there exist entangled states that do not violate Bell inequalities \cite{Werner89, Barrett02}. More recently, it has also been shown that there exist incompatible measurements that are useless for a Bell violation \cite{BV18, HQB18}. While the relationship between entanglement and Bell nonlocality has been an area of active research for over three decades \cite{Werner89, Barrett02, MLD08, HQB13, HQV16, CGR16}, in recent years research on the relationship between measurement incompatibility and Bell nonlocality has also picked up, often via the intermediate notion of quantum steering \cite{WPF09, QVB14, UMG14, BV18, HQB18, UBG15, KHC22, CBL16, QBW19}.

A quantum measurement is most generally represented by a positive operator-valued measure (POVM) \cite{HRS08}. A set of measurements is said to be jointly measurable (or compatible) if there exists a single measurement that can be coarse-grained in different ways to realize each of the measurements in the set \cite{HRS08}. Otherwise, it is said to be incompatible (\textit{i.e.,} not jointly measurable). In general, a set of measurements will exhibit various incompatibility relations among the measurements, \textit{e.g.,} a set of two measurements may either be compatible or incompatible while a set of three measurements can have a more complicated set of incompatibility relations such as pairwise compatibility but triplewise incompatibility \cite{LSW11}.\footnote{While commutativity is equivalent to joint measurability in the case of projective measurements, it is only sufficient (and not necessary) for joint measurability of general measurements \cite{HRS08}. A joint measurability structure that exhibits pairwise compatibility but triplewise incompatibility---often called Specker's scenario \cite{LSW11,KG14,KHF14,GKS18,AK20} or hollow triangle \cite{HFR14,QVB14,UMG14,HQB18} ---is therefore \textit{only} possible with POVMs, not projective measurements.} Any particular pattern of such incompatibility relations can be represented by a hypergraph that we term a {\em joint measurability structure}:\footnote{This is also referred to as an \textit{abstract simplicial complex} \cite{KHF14}.} the vertices of this hypergraph represent measurements and its hyperedges represent all (and only) the jointly measurable subsets of vertices.

To ask whether incompatibility is sufficient for Bell nonlocality is to ask the following question: given any set of incompatible POVMs that Alice can implement on her quantum system, does there always exist a bipartite entangled state that she can share with Bob and a set of POVMs that Bob can implement on his part of the state such that their joint statistics violates some Bell inequality? This question was recently settled in the negative via explicit counter-examples \cite{BV18,HQB18}. One of the counter-examples consists of three measurements that are pairwise compatible but triplewise incompatible \cite{BV18} (forming Specker's scenario \cite{Specker60, Specker60_1, Specker60_2, LSW11,KG14}) and the other involves an uncountably infinite number of measurements \cite{HQB18}. Thus, measurement incompatibility does not imply Bell nonlocality, similar to how there exist entangled states that do not violate Bell inequalities \cite{Werner89,Barrett02}.

It is interesting to note that although Specker's scenario admits a set of POVMs that do not violate any Bell inequality, it also admits sets of POVMs that do violate a Bell inequality \cite{QVB14,BV18}. This leads us to the following line of inquiry: The necessity of incompatibility for Bell nonlocality implies the necessity of a \textit{non-trivial joint measurability structure} on each wing of a Bell experiment, \textit{i.e.,} the joint measurability structure of the measurements on each wing should contain a subset of incompatible vertices. Is a non-trivial joint measurability structure, however, sufficient for a Bell inequality violation? 

That is, despite the inequivalence of incompatibility and Bell nonlocality in a \textit{quantitative} sense \cite{BV18,HQB18}, we ask whether a \textit{qualitative} equivalence between incompatibility and Bell nonlocality nevertheless holds: given any non-trivial joint measurability structure (with, say, $v$ vertices), can Alice implement a set of $v$ POVMs satisfying it on her part of some entangled state shared with Bob such that for some set of POVMs implemented by Bob on his part of this state their joint statistics violates a Bell inequality? Note that we want the number of settings for Alice in the Bell scenario that contribute to the Bell inequality violation---in the sense that the Bell inequality violation is a non-constant function of the outcome probabilities associated with each such setting---to be equal to the number of vertices in the joint measurability structure.
This is an important point because any joint measurability structure exhibiting some incompatibility necessarily admits a subset of incompatible POVMs and, as such, can be used to demonstrate the violation of a Bell inequality if this subset is a set of $N$-Specker POVMs (where $N\geq 2$, \textit{cf.}~Definition \ref{def:nspecker} and Refs.~\cite{CHSH, WPF09, BV18}) and if Alice chooses to implement only this particular subset of measurements and ignores the rest of the measurements in the joint measurability structure.
Such a violation would not be a property of the intended joint measurability structure but only a sub-structure thereof. 

In particular, arguments from input-lifting of Bell inequalities \cite{Pironio05} do not suffice to answer the question we raise here. Briefly, “input-lifting” refers to the fact that a facet Bell inequality in a Bell scenario remains a facet Bell inequality even if extra inputs are added to define an extended Bell scenario. Hence, a violation of the Bell inequality in the original scenario—viewed as a restriction of the extended scenario where the extra inputs do not contribute to the Bell expression—is a witness of Bell nonlocality for the extended scenario. A simple way to see why input-lifting doesn't work for our purposes is the following: we require that 1) any extra settings that are added on Alice’s wing in the original Bell scenario must respect incompatibility relations dictated by Alice’s particular joint measurability structure, and 2) that these settings contribute to the Bell expression involved in the Bell inequality violation. One cannot, in the absence of these two requirements, appeal merely to input-lifting to claim that a Bell inequality violation achieved by an $N$-Specker scenario embedded inside a joint measurability structure with $v$ vertices ($v>N$) is sufficient to conclude that the same violation is also a genuine property of the full joint measurability structure (rather than merely a sub-structure thereof). This would, for example, permit
a situation where the $v-N$ extra settings might as well be trivial, \textit{e.g.}, they always yield a fixed measurement outcome with certainty (thus being classical and compatible with all measurements) and do not, therefore, respect the $v$-vertex joint measurability structure whose ability to violate Bell inequalities we want to assess in the first place. To sum up, input-lifting does not on its own ensure that the full joint measurability structure will be respected on Alice's wing and, by definition, it ensures that the statistics of any extra inputs are irrelevant to the Bell inequality (violation) being lifted to the extended scenario.

We will answer our question in the affirmative by providing an explicit recipe for constructing a quantum realization of any joint measurability structure that also leads to a Bell inequality violation.

\textit{Preliminaries.---}We now formally define the basic notions we will use in the rest of this paper.

 Our definitions are equivalent to those in Refs.~\cite{KHF14,AK20}.
\begin{definition}[Positive operator-valued measure (POVM)]  
A general quantum measurement $M$ on a Hilbert space $\mathcal{H}$ is described by a set of positive semidefinite operators $\{M_a|M_a:\mathcal{H}\rightarrow\mathcal{H}, M_a\geq 0\}_{a\in\mathcal{O}}$, such that $\sum_{a\in\mathcal{O}} M_a=\id$, where $\id$ denotes the identity operator on $\mathcal{H}$ and $\mathcal{O}$ denotes the set of measurement outcomes.
\end{definition}

\begin{definition}[Joint measurability]\label{jmdef} A set of POVMs $\mathcal{M}=\{M_x\}_{x=1}^N$, each with outcome set $\mathcal{O}_x$, is said to be jointly measurable or compatible if it admits a (joint) POVM $G$, with outcome set $\mathcal{O}:=\mathcal{O}_1\times\mathcal{O}_2\times\dots\times\mathcal{O}_N$, such that each POVM $M_x:=\{M_{a_x|x}\}_{a_x\in\mathcal{O}_x}\in\mathcal{M}$ can be obtained as coarse-graining of $G$ over the outcomes of all other POVMs in $\mathcal{M}\backslash\{M_x\}$, \textit{i.e.,} $M_{a|x}=\sum_{\vec{a}\in\mathcal{O}}^{a_x=a}G(\vec{a})$, for all $a_x=a\in \mathcal{O}_x$, $x\in\{1,2,\dots,N\}$. Here $\vec{a}=(a_x)_{x=1}^N\in \mathcal{O}$.
\end{definition}
A set of POVMs that is not compatible is said to be \textit{incompatible}. Given a set of POVMs, its different subsets may or may not be compatible and such (in)compatibility relations can be expressed via their joint measurability structure.
\begin{definition}[Joint measurability structure]\label{jmsdef} A joint measurability structure on a set of POVMs $\mathcal{M}$ is a hypergraph $(V_\mathcal{M},E_\mathcal{M})$, with the set of vertices $V_\mathcal{M}$, each vertex representing a POVM in $\mathcal{M}$, and a set of hyperedges $E_\mathcal{M}=\{e|e\subseteq V^\mathcal{M}\}$ denoting all and only compatible (or jointly measurable) subsets of $\mathcal{M}$. Since every subset of a compatible set of POVMs is also compatible, in a valid joint measurability structure we must have $e'\subset e\in E_\mathcal{M}\Rightarrow e'\in E_\mathcal{M}$.
\end{definition}
A trivial joint measurability structure is one where all the vertices represent a compatible set of POVMs, \textit{i.e.}, it has exactly one hyperedge containing all the vertices. Any  joint measurability structure that is not trivial is said to be \textit{non-trivial}. A joint measurability structure is said to be \textit{quantum-realizable} if and only if there exist quantum measurements that can be assigned to its vertices such that these measurements satisfy all the (in)compatibility relations specified by the joint measurability structure. In Ref.~\cite{KHF14} it was shown that all joint measurability structures admit quantum realizations via an explicit construction. Crucial to this construction is a particular class of joint measurability structures called $N$-Specker scenarios.
\begin{definition}[$N$-Specker scenario]\label{def:nspecker} An $N$-Specker scenario is a joint measurability structure on a set of $N\geq2$ incompatible measurements where every $(N-1)$-element subset of the set is compatible. 
\end{definition}
Note that a $2$-Specker scenario corresponds to a pair of incompatible measurements, a $3$-Specker scenario corresponds to the situation Specker originally considered \cite{Specker60, Specker60_1, Specker60_2, LSW11, KG14}, and more general $N$-Specker scenarios have also been studied in the literature \cite{ULMH16,AK20}.

\begin{definition}[Bell-violating quantum realization of a joint measurability structure] A quantum realization of a joint measurability structure with $v$ vertices is said to be Bell-violating if and only if, given that a party (say, Alice) locally implements all the measurements from this realization, there exists an entangled state on a quantum system she can share with another party (say, Bob) and there exist local measurements on the other party's share of the state such that the entangled state subjected to their local measurements violates a Bell inequality in a Bell scenario with $v$ non-redundant settings for Alice, \textit{i.e.}, all the $v$ measurements from Alice's quantum realization contribute to the Bell violation.
\end{definition}

\textit{Resolution of any non-trivial joint measurability structure into $N$-Specker scenarios.---}We recall now the informal argument for this resolution first presented in Ref.~\cite{KHF14} before going on to provide a rigorous proof based on the properties of partially ordered sets in Appendix \ref{app:A}.\footnote{While Ref.~\cite{KHF14} left the argument for this resolution implicit in its constructive proof of quantum realizations of arbitrary joint measurability structures, in our proof we make it explicit that the underlying argument for such hypergraphs or abstract simplicial complexes \cite{KHF14} follows from the properties of partially ordered sets, \textit{i.e.}, independently of the question of quantum realizations of a joint measurability structure in the sense of Ref.~\cite{KHF14}.}

\begin{figure}
	\centering
	\includegraphics[scale=0.065]{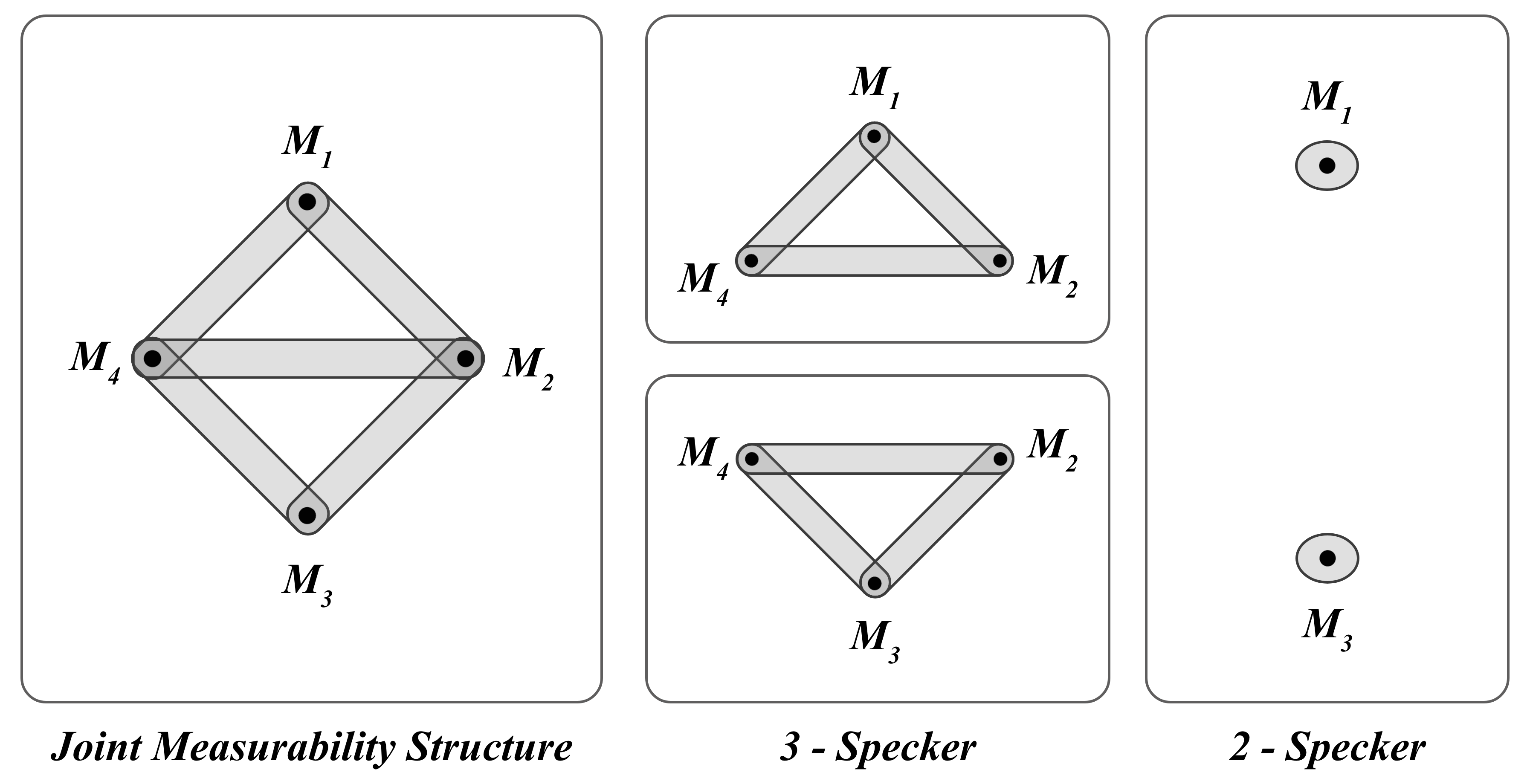}
	\caption{A joint measurability structure with $v=4$ and its decomposition into $3$-Specker and $2$-Specker scenarios.} \label{examplejmstruct}
\end{figure}

Consider any joint measurability structure $\mathcal{J}$ with $v$ vertices, where $V(\mathcal{J})$ denotes the set of vertices and $E(\mathcal{J})$ denotes the set of hyperedges, so that the cardinality of the set $V(\mathcal{J})$ is given by $|V(\mathcal{J})|=v$. We resolve $\mathcal{J}$ into $N$-Specker scenarios embedded in the hypergraph, where $N\in\{2,3,\dots,N_{\rm max}\}$ and $N_{\rm max}\leq v$, following the  method proposed in Ref.~\cite{KHF14}.\footnote{Note that $v=N_{\max}$ is the special case where $\mathcal{J}$ is a $v$-Specker scenario, \textit{i.e.}, every proper subset of $V(\mathcal{J})$ is compatible. In this case, the incompatibility of $\mathcal{J}$ cannot be ``localized" or ``reduced" to that of some strict subsets of $V(\mathcal{J})$.} This resolution of $\mathcal{J}$ into $N$-Specker scenarios rests on the following observation: an $N$-Specker scenario ($N\geq 2$) is a minimal incompatible set, \textit{i.e.}, any proper subset of it is compatible, and  thus serves as an irreducible unit of incompatibility in the joint measurability structure. As such, the incompatibility relations among the $v$ vertices of $\mathcal{J}$ can be fully captured by identifying these irreducible units of incompatibility in $\mathcal{J}$. We use the joint measurability structure in Fig.~\ref{examplejmstruct} as our working example to illustrate how this works: the incompatible subsets in this case are given by
\begin{align}\label{incompsubsets}
\{&\{M_1,M_2,M_3,M_4\},\{M_1,M_2,M_4\},\{M_2,M_3,M_4\},\nonumber\\
&\{M_1,M_3,M_4\},\{M_1,M_3,M_2\},\{M_1,M_3\}\}.
\end{align}
 
Of these incompatible subsets, the minimal ones are
\begin{align}
	\{\{M_1,M_2,M_4\},\{M_2,M_3,M_4\},\{M_1,M_3\}\}.
\end{align}

The non-minimal incompatible subsets in $\mathcal{J}$ can be generated from the minimal ones by progressively adding a new vertex to each minimal subset until the full set of vertices in $\mathcal{J}$ is covered: \textit{e.g.}, the minimal set $\{M_1,M_3\}$ generates the non-minimal sets $\{\{M_1,M_3,M_4\},\{M_1,M_3,M_2\}, \{M_1,M_2,M_3,M_4\}\}$; the other two minimal sets generate the non-minimal set $\{M_1,M_2,M_3,M_4\}$. Taking the union of these incompatible sets generated from the minimal ones yields the full family of incompatible subsets (Eq.~\eqref{incompsubsets}) in the joint measurability structure of Fig.~\ref{examplejmstruct}. The same argument for resolution into $N$-Specker scenarios generalizes to any $\mathcal{J}$. A formal proof of this resolution into $N$-Specker scenarios for any $\mathcal{J}$ can be found in Appendix \ref{app:B}.

We label the set of all $N$-Specker scenarios in the resolution of a joint measurability structure $\mathcal{J}$ by $\{{\rm Sp}_s(\mathcal{J})\}_s$, $s$ denoting a particular scenario in the resolution.

\textit{Construction of Bell-violating quantum realization of any non-trivial joint measurability structure.---}We prove our main result in two steps. First, we argue that every $N$-Specker scenario ($N\geq 2$) admits a violation of the $I_{NN22}$ Bell inequality \cite{CG04}, \textit{i.e.}, in a bipartite Bell scenario where each party has access to $N$ dichotomic POVMs. Second, we build a quantum realization for any joint measurability structure with $v$ vertices violating the $I_{vv22}$ Bell inequality.

The fact that every $N$-Specker scenario ($N\geq 2$) admits a Bell-violating quantum realization follows from a combination of the results of Refs.~\cite{CHSH, BV18}. In particular, Ref.~\cite{BV18} constructs a Bell-violating quantum realization for any $N$-Specker scenario where $N\geq 3$, violating the $I_{NN22}$ Bell inequality. On the other hand, we know that any $2$-Specker scenario can be realized by a pair of incompatible POVMs and, as such, admits a Bell-violating quantum realization, \textit{e.g.}, via the CHSH game \cite{CHSH, BCP14} with $X$ and $Z$ measurements for Alice. In fact, a stronger statement can be made in the $2$-Specker case (which doesn't hold for $N\geq 3$): specifically, Ref.~\cite{WPF09} shows that every quantum realization of a $2$-Specker scenario (\textit{i.e.}, every pair of incompatible POVMs) violates the CHSH inequality \cite{CHSH, BCP14}, which corresponds to the $I_{2222}$ Bell inequality. Hence, we have that every $N$-Specker scenario ($N\geq 2$) admits a quantum realization violating the $I_{NN22}$ Bell inequality.

We can now use these Bell-violating quantum realizations of $N$-Specker scenarios to construct a Bell-violating quantum realization of any non-trivial $v$-vertex joint measurability structure $\mathcal{J}$ that admits the resolution $\{{\rm Sp}_s(\mathcal{J})\}_s$ into $N$-Specker scenarios for $2\leq N\leq v$. We label the measurement settings corresponding to the Bell-violating realization of $\mathcal{J}$ by $x,y\in\{0,1,2,\dots,v-1\}$, respectively, for Alice and Bob. However, when considering a Bell-violating quantum realization of a particular $N$-Specker scenario ${\rm Sp}_s(\mathcal{J})$, for notational convenience we will temporarily relabel these $N$ measurement settings of Alice and Bob given by $x,y\in\{0,1,\dots,v-1\}$ (respectively) to $x',y'\in\{1,2,\dots,N-1\}$.

Consider a Bell-violating quantum realization of some ${\rm Sp}_s(\mathcal{J})$. We denote the entangled state used in the Bell-violating quantum realization of any ${\rm Sp}_s(\mathcal{J})$ as $\rho_s\in \mathcal{B}(\mathcal{H}_s\otimes\mathcal{H}_s)$, the POVMs of Alice as $\{M^{(s)}_{0|x'},M^{(s)}_{1|x'}\}$, and the POVMs of Bob as $\{M^{(s)}_{0|y'},M^{(s)}_{1|y'}\}$. For each $N$-Specker scenario ${\rm Sp}_s(\mathcal{J})$, the statistics $p(a,b|x',y')=\Tr(\rho_s M^{(s)}_{a|x'}\otimes M^{(s)}_{b|y'})$ (where $a,b\in\{0,1\}$) violates the $I^{(s)}_{NN22}$ Bell inequality, where $N=|V({\rm Sp}_s(\mathcal{J}))|$, the number of vertices in the scenario.  This Bell inequality \cite{CG04, BV18} is given by:
\begin{align}
	&I^{(s)}_{NN22}\nonumber\\
	:=& -p^{(s)}_B(0|1)-\sum_{x'=2}^{N}p^{(s)}_A(0|x')+\sum_{x'=1}^{N}p^{(s)}(00|x',y'=1)\nonumber\\
	&+\sum_{x'=2}^{N}p^{(s)}(00|x',x')-\sum_{1\leq x'<y' \leq N}p^{(s)}(00|x',y')
	\leq0.
\end{align}

Note that this realization does not assign POVMs to \textit{all} the $v$ vertices in $V(\mathcal{J})$. To the vertices that lie outside the $N$-Specker scenario ${\rm Sp}_s(\mathcal{J})$ under consideration---that is, $x\in V(\mathcal{J})\backslash V({\rm Sp}_s(\mathcal{J}))$---we assign the trivial POVMs $M^{(s)}_x=\{0,\id_s\}$, associating the outcome labelled ``0" with the impossible outcome and ``1" with the certain outcome. We do the same for Bob's measurements, adding measurements of the type $M^{(s)}_y=\{0,\id_s\}$ such that Bob now has $v$ POVMs, $v-N$ of them trivial. Now we have $v$ settings on each side of the Bell scenario but only $N$ of these on each side are non-trivial POVMs, the remaining $v-N$ settings being $\{0,\id_s\}$. We label the measurement settings by $x,y\in\{0,1,2,\dots,v-1\}$. The Hilbert space on which each $N$-Specker scenario ${\rm Sp}_s(\mathcal{J})$ is realized is $\mathcal{H}_s\cong \mathbb{C}^N$ (see Section I of the Supplemental Material). Since we know that the $N$-Specker scenario ${\rm Sp}_s(\mathcal{J})$ obtains the Bell inequality violation $I^{(s)}_{NN22}>0$, we also have that $I^{(s)}_{vv22}>0$. This is because the additional terms in the expression for $I^{(s)}_{vv22}$, beyond those coming from $I^{(s)}_{NN22}$, are all zero, as $p^{(s)}_A(0|x)=p^{(s)}_B(0|y)=0$ for all $x,y$ corresponding to trivial settings. Of course, $I^{(s)}_{vv22}>0$ is really just $I^{(s)}_{NN22}>0$ (since $v-N$ settings are trivial on each side), so it is not (yet) a Bell-violating realization of $\mathcal{J}$.

We can now combine the Bell violations from $N$-Specker scenarios (\textit{i.e.}, $I^{(s)}_{vv22}>0$) into a Bell-violating quantum realization of the given joint measurability structure $\mathcal{J}$, so that we will obtain $I_{vv22}>0$ in a bipartite Bell scenario with $v$ \textit{non-trivial} two-outcome measurement settings per party. Iterating our argument above for all $N$-Specker scenarios $\{{\rm Sp}_s(\mathcal{J})\}_s$ contained in $\mathcal{J}$, we get a set of states $\{\rho_s\}_s$ with measurements $\{M^{(s)}_x\}_{x=0}^{v-1}$ and $\{M^{(s)}_y\}_{y=0}^{v-1}$ associated with Alice and Bob, respectively. For each $s$, these states and measurements act on the Hilbert space $\mathcal{H}_s\otimes \mathcal{H}_s$ ($s$ runs through all the $N$-Specker scenarios in the decomposition of $\mathcal{J}$, \textit{cf.}~Fig.~\ref{examplejmstruct}). We now combine these constructions via a direct sum so that the Hilbert space on each side becomes $\mathcal{H}=\bigoplus_s \mathcal{H}_s$ and we have states and measurements defined on the tensor product space $\mathcal{H}\otimes\mathcal{H}\equiv(\bigoplus_s\mathcal{H}_s)\otimes(\bigoplus_s\mathcal{H}_s)$ in such a way that they violate the $I_{vv22}$ Bell inequality, \textit{i.e.}, $I_{vv22}>0$.
The joint measurability structure on Alice's side obtained via this construction is exactly the one given by $\mathcal{J}$: for every compatible subset $e\in E(\mathcal{J})$, its vertices are assigned POVMs that are compatible on every $\mathcal{H}_s$, and for every incompatible subset $e'\subseteq V(\mathcal{J})$, its vertices are assigned POVMs that are incompatible on some $\mathcal{H}_s$. 

The Bell-violating state and measurements on $\mathcal{H}\otimes\mathcal{H}$ are given as follows: we define $\rho:=[\bigoplus_{s_*} r_{s_*} \rho_{s_*}]\oplus [\bigoplus_{s\neq s'} 0_{s,s'}]\in\mathcal{B}(\mathcal{H}\otimes\mathcal{H})\cong \bigoplus_{s_*}\mathcal{B}(\mathcal{H}_{s_*}\otimes\mathcal{H}_{s_*})\oplus \bigoplus_{s\neq s'}\mathcal{B}(\mathcal{H}_{s}\otimes\mathcal{H}_{s'})$, where $r_{s_*}={\rm dim}(\mathcal{H}_{s_*})/{\rm dim}(\mathcal{H})$, $0_{s,s'}$ being the null operator on $\mathcal{H}_s\otimes \mathcal{H}_{s'}$, and measurements $M_{0|x}:=\bigoplus_s M^{(s)}_{0|x}, M_{0|y}:=\bigoplus_s M^{(s)}_{0|y}$. Hence, $\rho$ is restricted to acting non-trivially only on the subspace $\bigoplus_{s_*}(\mathcal{H}_{s_*}\otimes \mathcal{H}_{s_*})$ of $\mathcal{H}\otimes \mathcal{H}$ and its components outside of this subspace are zero. All the indices $s_*$, $s$, $s'$ run over the $N$-Specker scenarios $\{{\rm Sp}_s(\mathcal{J})\}_s$ contained in $\mathcal{J}$. It is easy to verify that this setup defined on $\mathcal{B}(\mathcal{H}\otimes\mathcal{H})$ violates a Bell inequality: specifically, noting that $x,y\in\{0,1,\dots,v-1\}$, we show $I_{vv22}>0$ for this choice of shared state and local measurements in Appendix \ref{app:B}. In fact, our construction does something more: namely, it also yields a Bell-violating quantum realization for \textit{each} non-trivial joint measurability structure contained in the given joint measurability structure, \textit{i.e.}, \textit{each} joint measurability structure formed by some incompatible subset of vertices (see Appendix \ref{app:C}). We therefore have:
\begin{theorem}\label{thm:qualequiv}
Every non-trivial joint measurability structure over a finite set of vertices admits a Bell-violating quantum realization that is also Bell-violating	for every non-trivial joint measurability structure contained in it.
\end{theorem}
Theorem \ref{thm:qualequiv} answers, in particular, the following question posed in Ref.~\cite{QVB14} in the affirmative:
\begin{quotation}
	Considering a set of arbitrarily many POVMs, it is known that any partial compatibility configuration can be realized [33]. Is it then possible to violate a Bell inequality for any possible configuration?
\end{quotation}
The ``partial compatibility configuration" of Ref.~\cite{QVB14} are the same as the non-trivial joint measurability structures we consider here \cite{KHF14}. 

Having thus shown that every non-trivial joint measurability structure admits a Bell-violating quantum realization, the following natural question arises: 

\textit{Given a non-trivial joint measurability structure, which quantum realizations of it are Bell-violating and which ones are not?}

Although we do not have a complete answer to this question for any given joint measurability structure, we studied it in the simplest case of interest, namely, the $3$-Specker scenario. We refer to the Supplemental Material (Section III) for further details and simply note here that our study led us to a new family of qubit POVMs that constitute Bell-violating quantum realizations of the 3-Specker scenario, substantially generalizing a planar family of POVMs proposed in Ref.~\cite{QVB14}.

\textit{Discussion.---}We have demonstrated a qualitative equivalence between incompatibility and Bell nonlocality, \textit{i.e.}, a non-trivial joint measurability structure is not only necessary but also \textit{sufficient} for Bell nonlocality. This sheds new light on the conceptual relationship between these two nonclassical features of quantum theory and raises interesting questions for future work. We already know that the mere fact of incompatibility of a set of measurements is not enough to imply a Bell inequality violation \cite{BV18, HQB18}. However, given the joint measurability structure of an incompatible set of measurements that cannot violate a Bell inequality, our result shows that we can always find another set of measurements with the same structure that do violate a Bell inequality. That is, there are no non-trivial joint measurability structures that are ``useless" for a Bell inequality violation. 

A fundamental question this raises is the following: given a quantum realization of a non-trivial joint measurability structure, which features of this realization are responsible for a Bell inequality violation? A characterization of this type would help us obtain a finer handle on the relationship between incompatibility and Bell nonlocality by allowing us to target, when required, those measurements that are useful for Bell inequality violations. While the general characterization problem may be difficult to solve for arbitrary joint measurability structures, a lot of insight can be gained by studying Bell-violating vs.~Bell non-violating realizations of the simplest joint measurability structure beyond a pair of incompatible measurements, \textit{i.e.}, Specker's scenario with three binary outcome measurements. 

Although we have carried out some numerical investigations of the $3$-Specker scenario (see Supplemental Material), further analytical investigations will be essential for a deeper understanding of the interplay between Bell nonlocality and measurement incompatibility. For example, the family of qubit POVMs that constitute Bell-violating quantum realizations of the $3$-Specker scenario identified in the Supplemental Material generalizes the planar family of POVMs proposed in Ref.~\cite{QVB14}. Can we generalize this further, perhaps to non-planar measurements? Similarly, do the two families of qubit POVMs that do not show $I_{3322}$ violations (see Supplemental Materail) also fail to show Bell nonlocality more generally, going beyond known examples \cite{BV18,HQB18}? Finally, do the realizations of non-trivial joint measurability structures with qubit POVMs (including $N$-Specker scenarios for all finite $N$) obtained in Ref.~\cite{AK20} enable Bell inequality violations, perhaps on two-qubit systems? Such investigations will open up avenues for better leveraging the incompatibility of measurements in quantum protocols based on Bell nonlocality \cite{BCP14} and will be taken up in future work.

\ack
We would like to thank Máté Farkas, Tobias Fritz, Tomáš Gonda, and Paul Skrzypczyk for discussions and comments. This work received support from the French government under the France 2030 investment plan, as part of the Initiative d'Excellence d'Aix-Marseille Université-A*MIDEX, AMX-22-CEI-01. Part of this work was supported by the Chargé de Recherche fellowship of the Fonds de la Recherche Scientifique (F.R.S.-FNRS), Belgium and the Program of Concerted Research Actions (ARC) of the Université libre de Bruxelles. We also acknowledge support from the Perimeter Institute for Theoretical Physics through their Undergraduate Summer Research Program which led to this collaboration. Research at Perimeter Institute is supported by the Government of Canada through the Department of Innovation, Science and Economic Development Canada and by the Province of Ontario through the Ministry of Research, Innovation and Science.

\section*{Appendix}
\subsection{Every non-trivial joint measurability structure admits a resolution into $N$-Specker scenarios}\label{app:A}
We will use some generic properties of any partial order over a set in proving our claim. Namely, 
\begin{enumerate}
	\item Any partially ordered set, say $(X,\preceq)$, admits minimal elements.
	
	\item The upward closure of any element $x\in X$ is defined by the (upward closed) set $\uparrow x:=\{u| u\in X, x\preceq u\}$.\footnote{This set is ``upward closed" in the sense that it contains any element above $x$ in the partial order $(X,\preceq)$.}
	
	\item The union of the upward closures of the minimal elements of $(X,\preceq)$ is equal to  $X$, i.e., $$X=\bigcup_{{x}_{\rm min}}\uparrow x_{\rm min}.$$
	
	It is easy to see this as follows: Suppose some $x\in X$ but $x\notin \bigcup_{{x}_{\rm min}}\uparrow x_{\rm min}$. That is, $x$ is not above \textit{any} minimal element $x_{\rm min}$ in the partial order. Hence, for each minimal element $x_{\min}$, $x$ is either below $x_{\min}$ or incomparable to $x_{\min}$: the former is ruled out because nothing is below $x_{\min}$ by definition, and in the latter case, $x$ must be a minimal element itself and therefore cannot be outside the set $\bigcup_{{x}_{\rm min}}\uparrow x_{\rm min}$. Thus, our supposition is flawed and we have $X=\bigcup_{{x}_{\rm min}}\uparrow x_{\rm min}$.
\end{enumerate}

We now apply the above properties to the partial order over incompatible subsets in any given non-trivial joint measurability structure $\mathcal{J}$. Given $\mathcal{J}$, let us define the hypergraph, $\overline{\mathcal{J}}$, with vertices $V(\overline{\mathcal{J}})=V(\mathcal{J})$ and hyperedges $E(\overline{\mathcal{J}})=\{e\subseteq V(\mathcal{J})| e\notin E(\mathcal{J})\}$.\footnote{Note that both $\mathcal{J}$ and $\overline{\mathcal{J}}$ are valid representations of the (in)compatibility relations, the former denoting compatibility via hyperedges and the later denoting incompatibility. Further, the partially ordered set of interest here is $(X,\preceq):=(E(\overline{\mathcal{J}}),\subseteq)$, where the set $X$ denotes all incompatible subsets of $V(\overline{\mathcal{J}})$ and the partial order relation ($\preceq$) is given by set inclusion ($\subseteq$).} It is now easy to see how the resolution of $\mathcal{J}$ into $N$-Specker scenarios works using the joint measurability structure in Fig.~\ref{examplejmstruct} as our working example: the incompatible subsets (\textit{i.e.}, hyperedges of $\overline{\mathcal{J}}$) are given by
\begin{align}
	\{&\{M_1,M_2,M_3,M_4\},\{M_1,M_2,M_4\},\{M_2,M_3,M_4\},\nonumber\\
	&\{M_1,M_3,M_4\},\{M_1,M_3,M_2\},\{M_1,M_3\}\}.
\end{align}

Of these incompatible subsets, the minimal ones---namely, those for which every proper subset is compatible---are given by (\textit{cf.}, Fig.~\ref{examplejmstruct})
\begin{align}
	\{\{M_1,M_2,M_4\},\{M_2,M_3,M_4\},\{M_1,M_3\}\}.
\end{align}

The non-minimal incompatible sets in $\mathcal{J}$ (equivalently, hyperedges of $\overline{\mathcal{J}}$) can be generated from the minimal ones by progressively adding a new vertex to each minimal subset until the full set of vertices in $\mathcal{J}$ is covered: \textit{e.g.}, the minimal incompatible ($2$-Specker) set $\{M_1,M_3\}$ generates the non-minimal incompatible sets $\{\{M_1,M_3,M_4\},\{M_1,M_3,M_2\}, \{M_1,M_2,M_3,M_4\}\}$; the other two minimal incompatible sets generate the (non-minimal) incompatible set $\{M_1,M_2,M_3,M_4\}$. The same argument for resolution into $N$-Specker scenarios applies for any $\mathcal{J}$. The formal reason for this is that the incompatibility relations in $\mathcal{J}$ form a partial order with respect to set inclusion, \textit{i.e.}, for any two incompatible subsets $S_1,S_2$ of $V(\mathcal{J})$, $S_1\subseteq S_2$, $S_2\subseteq S_1$, or neither is a subset of the other, \textit{i.e.}, $S_1\nsubseteq S_2$ and $S_2\nsubseteq S_1$. The minimal elements of this partial order for any non-trivial joint measurability structure are given by $N$-Specker scenarios since they do not contain any proper subset that is incompatible, \textit{i.e.}, there are no incompatible subsets below them in this partial order over incompatible subsets. Similarly, the maximal element of any non-trivial joint measurability structure is the full set of vertices, $V(\mathcal{J})$. Denoting the minimal elements of this partial order---namely, $N$-Specker scenarios---by $\{{\rm Sp}_s(\mathcal{J})\}_s$, we have that
\begin{align}
	E(\overline{\mathcal{J}})&=\bigcup_s \uparrow E(\overline{{\rm Sp}_s(\mathcal{J})})\nonumber\\
	&=\bigcup_s \uparrow \{V({\rm Sp}_s(\mathcal{J}))\}
\end{align}
where $\uparrow E(\overline{{\rm Sp}_s(\mathcal{J})})$ denotes the upward-closure of $N$-Specker scenario ${\rm Sp}_s(\mathcal{J})$ under set inclusion. Here, $\overline{{\rm Sp}_s(\mathcal{J})}$ is the hypergraph obtained from ${\rm Sp}_s(\mathcal{J})$ via $V(\overline{{\rm Sp}_s(\mathcal{J})})=V({\rm Sp}_s(\mathcal{J}))$ and $E(\overline{{\rm Sp}_s(\mathcal{J})})=\{V({\rm Sp}_s(\mathcal{J}))\}$.

Hence, $\uparrow\{V({\rm Sp}_s(\mathcal{J}))\}$ is obtained by constructing incompatible subsets starting from the minimal incompatible subset $V({\rm Sp}_s(\mathcal{J}))$ and progressively adding other vertices from $V(\mathcal{J})$ to it, e.g., in Fig.~\ref{examplejmstruct}, the minimal incompatible subset $V({\rm Sp}_s(\mathcal{J}))=\{M_1,M_3\}$ has the upward-closure
\begin{align}
	&\uparrow\{\{M_1,M_3\}\}\nonumber\\
	=&\{\{M_1,M_3,M_4\},\{M_1,M_3,M_2\}, \{M_1,M_2,M_3,M_4\}\}	
\end{align}
under set inclusion.

\subsection{Proof that our quantum realization is Bell-violating for any non-trivial $\mathcal{J}$}\label{app:B}The statistics relevant for $I_{vv22}$ is determined by the following probabilities:
\begin{align}
	&p_A(0|x)=\Tr(\rho M_{0|x}\otimes \id)\nonumber\\
	=&\Tr\Big(\rho\big(\bigoplus_sM^{(s)}_{0|x}\big)\otimes \big(\bigoplus_{s'}\id_{s'}\big)\Big)\nonumber\\
	=&\Tr\Big(\rho\big(\bigoplus_s(M^{(s)}_{0|x}\otimes \id_s)\oplus \bigoplus_{s\neq s'}(M^{(s)}_{0|x}\otimes \id_{s'})\big)\Big)\nonumber\\
	=&\Tr\Big(\big(\bigoplus_{s_*} r_{s_*}\rho_{s_*}\big)\big(\bigoplus_s(M^{(s)}_{0|x}\otimes \id_s)\big)\Big)\nonumber\\
	=&\Tr\Big(\bigoplus_s r_s\rho_s(M^{(s)}_{0|x}\otimes \id_s)\Big)\nonumber\\
	=&\sum_sr_s\Tr(\rho_s M^{(s)}_{0|x}\otimes\id_s)\\
	=&\sum_sr_sp^{(s)}_A(0|x).
\end{align}
\begin{align}
	&p_B(0|y)=\Tr(\rho \id\otimes M_{0|y})\nonumber\\
	=&\Tr\Big(\rho\big(\bigoplus_{s'}\id_{s'}\big)\otimes\big(\bigoplus_{s}M^{(s)}_{0|y}\big)\Big)\nonumber\\
	=&\Tr\Big(\rho\big(\bigoplus_s( \id_s\otimes M^{(s)}_{0|y})\oplus\bigoplus_{s\neq s'}(\id_s\otimes M^{(s')}_{0|y})\big)\Big)\nonumber\\
	=&\Tr\Big(\big(\bigoplus_{s_*} r_{s_*}\rho_{s_*}\big)\big(\bigoplus_s(\id_s\otimes M^{(s)}_{0|y})\big)\Big)\nonumber\\
	=&\Tr\Big(\bigoplus_s r_s\rho_s\big(\id_s\otimes M^{(s)}_{0|y}\big)\Big)\nonumber\\
	=&\sum_sr_s\Tr(\rho_s \id_s\otimes M^{(s)}_{0|y})\\
	=&\sum_sr_sp^{(s)}_B(0|y).
\end{align}

\begin{align}
	&p(00|xy)=\Tr(\rho M_{0|x}\otimes M_{0|y})\nonumber\\
	=&\Tr\Big(\rho\big(\bigoplus_sM^{(s)}_{0|x}\big)\otimes \big(\bigoplus_{s'}M^{(s')}_{0|y}\big)\Big)\nonumber\\
	=&\Tr\Big(\rho\big(\bigoplus_s(M^{(s)}_{0|x}\otimes M^{(s)}_{0|y})\oplus\bigoplus_{s\neq s'}(M^{(s)}_{0|x}\otimes M^{(s')}_{0|y})\big)\Big)\nonumber\\
	=&\Tr\Big(\big(\bigoplus_{s_*} r_{s_*}\rho_{s_*}\big)\big(\bigoplus_s(M^{(s)}_{0|x}\otimes M^{(s)}_{0|y})\big)\Big)\nonumber\\
	=&\Tr\Big(\bigoplus_s r_s\rho_s\big(M^{(s)}_{0|x}\otimes M^{(s)}_{0|y}\big)\Big)\nonumber\\
	=&\sum_sr_s\Tr(\rho_s(M^{(s)}_{0|x}\otimes M^{(s)}_{0|y}))=\sum_sr_sp^{(s)}(00|xy).
\end{align}

We then have:
\begin{align}
	&I_{vv22}\nonumber\\
	\equiv&-p_B(0|0)-\sum_{x=1}^{v-1}p_A(0|x)+\sum_{x=0}^{v-1}p(00|x,y=0)\nonumber\\
	&+\sum_{x=1}^{v-1}p(00|x,x)-\sum_{0\leq x<y \leq v-1}p(00|x,y)\\
	=&\sum_sr_sI^{(s)}_{vv22}>0.
\end{align}
The last inequality follows from the fact that $I^{(s)}_{vv22}>0$ and $r_s>0$ for all ${\rm Sp}_s(\mathcal{J})$ in the resolution of $\mathcal{J}$. Hence, given any non-trivial joint measurability structure required of Alice's measurement settings, our construction provides a Bell-violating quantum realization of it.

\subsection{Proof that our quantum realization is Bell-violating for every non-trivial joint measurability structure contained in $\mathcal{J}$}\label{app:C}To see this for a non-trivial joint measurability structure, say $\mathcal{J}_{\rm sub}$, contained in $\mathcal{J}$, one simply has to note that a subset of $N$-Specker scenarios in the decomposition of $\mathcal{J}$ are sufficient to reconstruct $\mathcal{J}_{\rm sub}$, \textit{i.e.}, to obtain the full set of incompatibility relations in $\mathcal{J}_{\rm sub}$. Further, since $I^{(s)}_{vv22}>0$ and $r_s>0$ for all ${\rm Sp}_s(\mathcal{J})$ in the resolution of $\mathcal{J}$, this is also true for the ${\rm Sp}_s(\mathcal{J})$ in the resolution of $\mathcal{J}_{\rm sub}$. Denoting the number of vertices in $\mathcal{J}_{\rm sub}$ as $w$ $(\leq v)$, we have that

\begin{align}
	I_{ww22}:=&-p_B(0|0)-\sum_{x=1}^{w-1}p_A(0|x)+\sum_{x=0}^{w-1}p(00|x,y=0)\nonumber\\
	&+\sum_{x=1}^{w-1}p(00|x,x)-\sum_{0\leq x<y \leq w-1}p(00|x,y)\\
	=&\sum_{s: V(\mathcal{\rm Sp}_s(\mathcal{J}))\subseteq V(\mathcal{J}_{\rm sub})}r_sI^{(s)}_{ww22}>0,
\end{align}

where the sum is over those $N$-Specker scenarios $\mathcal{\rm Sp}_s(\mathcal{J})$ that are contained in $\mathcal{J}_{\rm sub}$.

\bibliographystyle{unsrturl}

	\newpage

	\title{Supplemental Material for `Qualitative equivalence between incompatibility and Bell nonlocality'}
	
	\author{Shiv Akshar Yadavalli}
	\email{sy215@duke.edu}
	\affiliation{Department of Physics, Duke University, Durham, North Carolina, USA 27708}
	
	\author{Nikola Andrejic}
	\email{nikola.andrejic@pmf.edu.rs}
	\affiliation{University of Ni\v s, Faculty of Sciences and Mathematics,  Vi\v segradska 33, 18000 Ni\v s, Serbia}
	
	\author{Ravi Kunjwal}
	\email{quaintum.research@gmail.com}
	\affiliation{Universit\'e libre de Bruxelles, QuIC, Brussels, Belgium}
	\affiliation{Aix-Marseille University, CNRS, LIS, Marseille, France}

\maketitle

	\section{Bell-violating quantum realization of ${\rm Sp}_s(\mathcal{J})$}\label{speckerbell} We now use the Bell-violating quantum realization of $N$-Specker scenarios in Ref.~\cite{BV18} to construct a quantum realization of each such scenario ${\rm Sp}_s(\mathcal{J})$ in the resolution $\{{\rm Sp}_s(\mathcal{J})\}_s$ of $\mathcal{J}$. That is, given a  ${\rm Sp}_s(\mathcal{J})$ in $\mathcal{J}$, we assign POVMs $\{M^{(s)}_x\}_{x=1}^N$ in $\mathcal{B}(\mathcal{H}_s)$ (where $\mathcal{H}_s\cong\mathbb{C}^N$) to the $N$ vertices of ${\rm Sp}_s(\mathcal{J})$ following the construction of Ref.~\cite{BV18}:
	\begin{align}
		M^{(s)}_{a=0|x}=\eta \ket{A_x}\bra{A_x}, \ket{A_x}=\sum^N_{j=1}A_{xj} \ket{j} 
	\end{align}
	where $\{\ket{j}\}_{j=1}^N$ is an orthonormal basis of $\mathbb{C}^N$, $\eta=1/(N-1)$, and the probability amplitudes $A_{xj}$ are defined as entries of the $N\times N$ matrix 
	\begin{equation}
		A=\begin{bmatrix}
			0&0&\dots&0&0&-q_1&-q_0\\
			0&0&\dots&0&-q_2&\frac{q_1}{N-1}&q_0\\
			0&0&\dots&-q_3&\frac{q_2}{N-2}&\frac{q_1}{N-1}&q_0\\
			\vdots&\vdots&\ddots\\
			-q_{N-1}&\frac{q_{N-2}}{2}&\dots&\frac{q_3}{N-3}&\frac{q_2}{N-2}&\frac{q_1}{N-1}&q_0\\
			q_{N-1}&\frac{q_{N-2}}{2}&\dots&\frac{q_3}{N-3}&\frac{q_2}{N-2}&\frac{q_1}{N-1}&q_0
		\end{bmatrix},
	\end{equation}
	where $q_1^2+q_0^2=1$ and $q_{k+1}^2=\left(1-\frac{1}{(N-k)^2}\right)q_k^2$ for $k\geq1$.
	This means that each row of $A$ denotes the unit state vector $\ket{A_x}$, \textit{i.e.}, $\sqrt{\sum_{j=1}^N|A_{xj}|^2}=1$ \cite{VPB10, BV18}. Hence, Alice has access to the POVMs $M^{(s)}_x\equiv\{M^{(s)}_{a=0|x},M^{(s)}_{a=1|x}=\id_s-M^{(s)}_{a=0|x}\}$ for all $x\in\{1,2,\dots,N\}$. Here, the condition that $\eta=\frac{1}{N-1}$ is sufficient to ensure that the joint measurability structure is an $N$-Specker scenario following the construction in Section 4 of Ref.~\cite{BV18}).
	
	To witness a Bell inequality violation, Bob is given access to $N$ POVMs $M^{(s)}_y\equiv\{M^{(s)}_{b=0|y},M^{(s)}_{b=1|y}=\id_s-M^{(s)}_{b=0|y}\}$ given by 
	\begin{align}
		M^{(s)}_{b=0|y}= \ket{B_y}\bra{B_y}, \ket{B_y}=\sum^N_{j=1}B_{yj} \ket{j} 
	\end{align}
	where $B_{yj}$ are entries of the $N\times N$ matrix
	\begin{equation}
		B=\begin{bmatrix}
			0&0&\dots&0&0&0&1\\
			0&0&\dots&0&-p_2&\frac{p_1}{N-1}&p_0\\
			0&0&\dots&-p_3&\frac{p_2}{N-2}&\frac{p_1}{N-1}&p_0\\
			\vdots&\vdots&\ddots\\
			-p_{N-1}&\frac{p_{N-2}}{2}&\dots&\frac{p_3}{N-3}&\frac{p_2}{N-2}&\frac{p_1}{N-1}&p_0\\
			p_{N-1}&\frac{p_{N-2}}{2}&\dots&\frac{p_3}{N-3}&\frac{p_2}{N-2}&\frac{p_1}{N-1}&p_0
		\end{bmatrix},
	\end{equation}
	where $p_0^2=\frac{1}{N}$, $p_1^2=\frac{N-1}{N}=1-\frac{1}{N}$, $p_{k+1}^2=\left(1-\frac{1}{(N-k)^2}\right)p_k^2$ for $k\geq1$. Again, each row of $B$ denotes the unit state vector $\ket{B_y}$, \textit{i.e.}, $\sqrt{\sum_{j=1}^N|B_{yj}|^2}=1$ \cite{VPB10,BV18}.
	
	The quantum state $\ket{\psi}_{\epsilon}\in\mathbb{C}^N\otimes\mathbb{C}^N$ shared between Alice and Bob is given by 
	
	\begin{equation}
		\ket{\psi_{\epsilon}}=\sqrt{\frac{1-\epsilon^2}{N-1}}\left(\sum_{k=1}^{N-1}\ket{k}\ket{k}\right)+\epsilon\ket{N}\ket{N}, \epsilon\in[0,1].
	\end{equation}
	
	The Bell scenario therefore consists of two parties, each with $N$ dichotomic measurements. The $I^{(s)}_{NN22}$ Bell inequality \cite{CG04} for this scenario is given by 
	\begin{align}\label{nspeckerbell}
		I^{(s)}_{NN22}:=& -p^{(s)}_B(0|1)-\sum_{x=2}^Np^{(s)}_A(0|x)+\sum_{x=1}^Np^{(s)}(00|x,y=1)\nonumber\\
		+&\sum_{x=2}^Np^{(s)}(00|x,x)-\sum_{1\leq x<y \leq N}p^{(s)}(00|x,y)\nonumber\\
		\leq& 0.
	\end{align} 
	Using the state and measurements above, we have (following Ref.~\cite{VPB10}) for the probabilites entering $I^{(s)}_{NN22}$:
	\begin{align}
		p^{(s)}_B(0|1)&=\epsilon^2,\\
		p^{(s)}_A(0|x)&=\frac{1-\epsilon^2}{N-1}(1-q_0^2)+\epsilon^2q_0^2\nonumber\\
		&\textrm{ for }2\leq x\leq N,\\
		p^{(s)}(00|x,y=1)&=\epsilon^2q_0^2\nonumber\\
		&\textrm{ for }1\leq x\leq N,\\
		p^{(s)}(00|x,x)&=\left(\sqrt{\frac{1-\epsilon^2}{N-1}}p_1q_1+\epsilon p_0q_0\right)^2\nonumber\\ &\textrm{ for }2\leq x\leq N,\\
		p^{(s)}(00|x,y)&=\left(\sqrt{\frac{1-\epsilon^2}{N-1}}
		\frac{p_1q_1}{1-N}+\epsilon p_0q_0\right)^2\nonumber\\
		&\textrm{ for }1\leq x<y\leq N.
	\end{align}
	Following Ref.~\cite{VPB10}, we choose $\epsilon$ to maximize the violation of Eq.~\eqref{nspeckerbell}, i.e.,
	\begin{equation}
		\epsilon^2=\frac{1-q_0^2}{1+[(N-1)^2-1]q_0^2},
	\end{equation}
	so that the terms $p^{(s)}(00|x,y)$ are set to zero for all $x,y$ such that $1\leq x<y\leq N$. We then have that 
	\begin{equation}
		I^{(s)}_{NN22}=\eta\epsilon^2\left(-\frac{1}{\eta}+q_0^2N\right)>0, \textrm{ for }\eta>\frac{1}{Nq_0^2},
	\end{equation}
	and $\eta=\frac{1}{N-1}>\sqrt{\frac{1}{Nq_0^2}}$ requires that $q_0>\left(1-\frac{1}{N}\right)$. Hence, $q_0>\left(1-\frac{1}{N}\right)$ implies that $\eta=\frac{1}{N-1}$ is sufficient to violate the inequality $I^{(s)}_{NN22}\leq 0$.
	
	\section{Bell violation from any $2$-Specker scenario}
	We consider any ${\rm Sp}_s(\mathcal{J})$ in the resolution of $\mathcal{J}$ that is just a pair of incompatible vertices. Under the no-signalling condition, the expression for $I^{(s)}_{NN22}$ Bell inequality reduces to the CHSH inequality for $N=2$. That is, the inequality (also, referred to as the CH inequality \cite{CH74})
	\begin{align}
		&I^{(s)}_{2222}\nonumber\\
		=&-p^{(s)}_B(0|1)-p^{(s)}_A(0|2)+p^{(s)}(00|1,1)+p^{(s)}(00|2,1)\nonumber\\
		+&p^{(s)}(00|2,2)-p^{(s)}(00|1,2)\\
		&\leq 0
	\end{align}
	is equivalent to the Bell-CHSH inequality \cite{CHSH}.
	We refer the reader to Ref.~\cite{CH74} for a proof of this equivalence (see also \cite{Cereceda01} for a more modern treatment).
	
	Now, to obtain a violation of the CHSH inequality from a $2$-Specker scenario, all we need is any pair of incompatible dichotomic POVMs. Wolf {\em et al.}~\cite{WPF09} showed that Alice can use any such pair POVMs to violate the CHSH inequality for some choice of entangled state shared with Bob and some choice of POVMs for Bob. For example, one could take the shared state $\rho_s$ to be a two-qubit maximally entangled state (hence $\mathcal{H}_s\cong \mathbb{C}^2$) and the measurements $\{M^{(s)}_x\}_{x=1}^2$ and $\{M^{(s)}_y\}_{y=1}^2$ to be those that achieve Tsirelson's bound \cite{BCP14}. The assignment of trivial $\{0,\id_s\}$ POVMs to the rest of the $v-2$ vertices in $\mathcal{J}$ can proceed as we did in the $N\geq 3$ case.
	
	\section{Bell violations in the $3$-Specker scenario}\label{app:3specker}
	We investigate the ability of three qubit POVMs with coplanar Bloch vectors to violate the $I_{3322}$ Bell inequality \cite{CG04} in the $3$-Specker scenario. In Ref.~\cite{BV18} it was shown that following set of trine spin POVMs form a $3$-Specker scenario but never violate a Bell inequality (thus showing that incompatibility does not imply Bell nonlocality):
	\begin{align}\label{bvexample}
		\forall k\in\{1,2,3\},\hspace{5pt} &x_k\in\{-1,+1\}:\nonumber\\ E_k(x_k)&=\frac{1}{2}\left(I+0.67x_k\vec{n}_k\cdot\vec{\sigma}\right),\nonumber\\
		\textrm{where }\vec{n}_k&=\cos\frac{2k\pi}{3} \vec{e}_x + \sin\frac{2k\pi}{3} \vec{e}_z,\nonumber\\
		\textrm{and }\vec{e}_x&=(1,0,0), \vec{e}_z=(0,0,1), \vec{\sigma}=(\sigma_x,\sigma_y,\sigma_z).
	\end{align}
	This inspires us to look at two families of POVMs to which the set in Eq.~\eqref{bvexample} belongs, \textit{i.e.}, to two generalizations of Eq.~\eqref{bvexample}. In both of these generalizations we keep the trine directions but we introduce a global purity parameter $\eta\in[0,1]$ and a bias $b\in[-1,1]$ such that $|b|\leq 1-\eta$. Since bias introduces a preferred orientation along the Bloch line there are two ways to assign biases producing the two families of interest, \textit{i.e.},
	\begin{subequations}
		\begin{align}\label{fam1}
			E_k(x_k)&=\frac{1}{2}\left((1+x_k b)I+\eta x_k\vec{n}_k\cdot\vec{\sigma}\right),\\
			E_k(x_k)&=\frac{1}{2}\left((1+(-1)^k x_k b)I+\eta x_k\vec{n}_k\cdot\vec{\sigma}\right),\label{fam2}\\
			\textrm{where }&\vec{n}_k = \cos\frac{2k\pi}{3} \vec{e}_x + \sin\frac{2k\pi}{3} \vec{e}_z,\nonumber\\
			&k\in\{1,2,3\}, \hspace{5pt} x_k\in\{-1,+1\}.
		\end{align}
	\end{subequations}
	In contrast to the trine example used in Ref.~\cite{BV18} (namely, Eq.~\eqref{bvexample}), Ref.~\cite{QVB14} provides an example of POVMs with trine spin directions that does form a $3$-Specker scenario and violates Bell's inequalities. Generalizing the construction of Ref.~\cite{QVB14}, we define another family of POVMs as 
	\begin{subequations}
		\begin{align}
			E_1(+1) &= \frac{1}{2}\alpha\left(I+\vec{n}_1\cdot\vec{\sigma}\right),\\
			E_2(+1) &= \frac{1}{2}\alpha\left(I+\vec{n}_2\cdot\vec{\sigma}\right),\\
			E_3(+1) &= \frac{1}{2}\beta\left(I-\vec{n}_3\cdot\vec{\sigma}\right),
		\end{align}\label{quintino}
	\end{subequations}
	which recovers the example from \cite{QVB14} when $\beta=3\alpha/5$, where $\alpha,\beta\in[0,1]$.
	
	For the three families of trine POVMs we have defined (Eqs.~\eqref{fam1}, \eqref{fam2},  \eqref{quintino}), we will investigate the dependence of the joint measurability structure they realize and their ability to violate the $I_{3322}$ Bell inequality on the parameters we have introduced in each case. For each family, the pairwise joint measurability is determined analytically \cite{YLL10}, while the triplewise joint measurability condition is obtained by solving a semidefinite program (SDP) that maximizes that relevant parameters in the POVMs subject to the existence of a triplewise joint POVM.
	
\begin{figure*}
		\centering
		\includegraphics[scale=0.2]{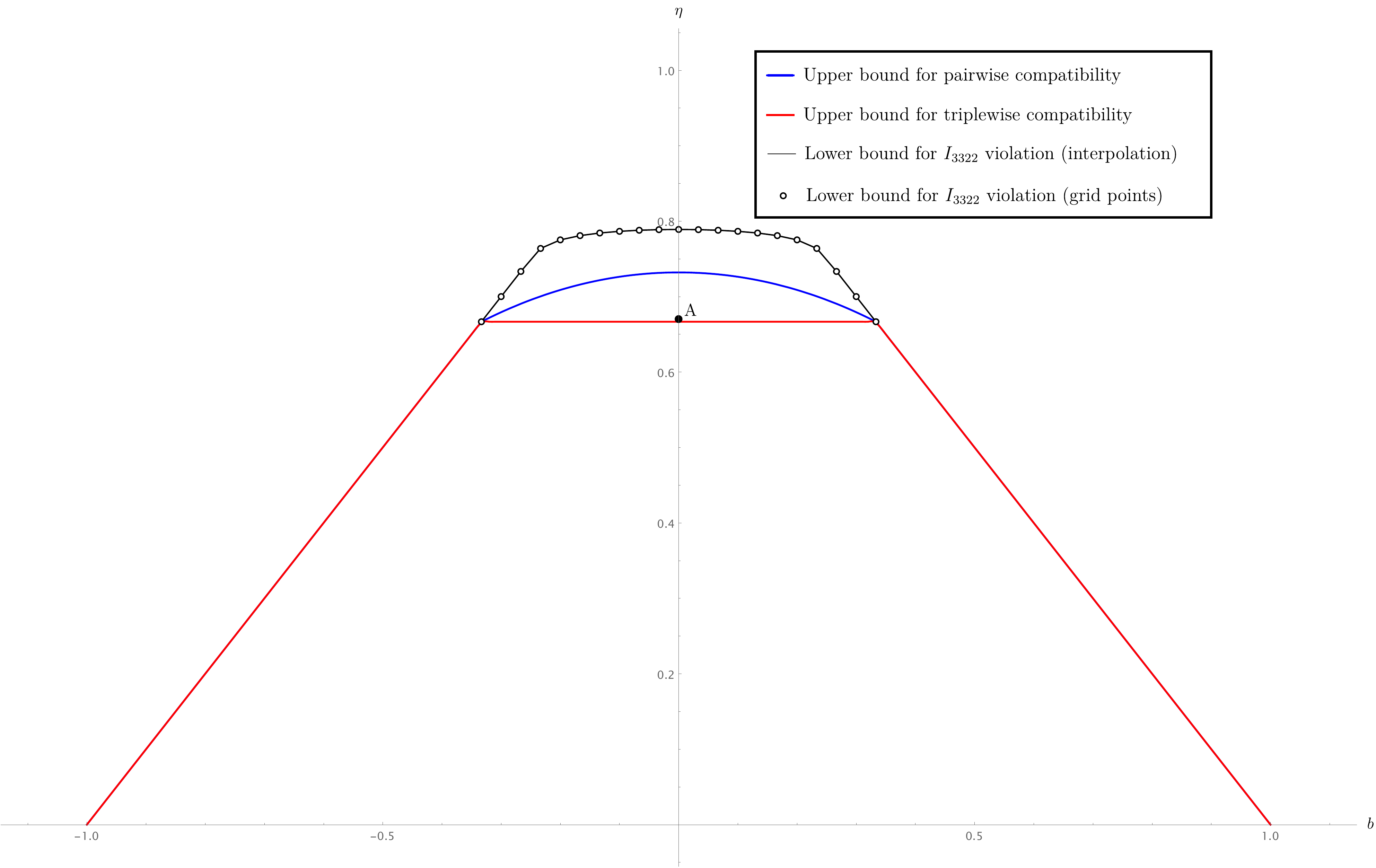}
		\caption{Red line represents the upper bound for $\eta$ given $b$ such that the whole set is compatible. Blue line (coinciding with the red one for $|b|>1/3$) represents the upper bound for $\eta$ such that each pair is jointly measurable. Black line represents lower bound for $\eta$ such that we have $I_{3322}$ violation. The points along the black line are numerically obtained and the rest of it is an extrapolation. In the region between the blue and the red line, where we have a 3-Specker scenario, there are no $I_{3322}$ violations. Point $A$ represents the  POVMs used for Alice (same as our Eq.~\eqref{bvexample}) in Ref.~\cite{BV18} to show that measurement incompatibility does not imply Bell nonlocality.}
		\label{fig:bias1}
		
		\bigskip
		
		\includegraphics[scale=0.25]{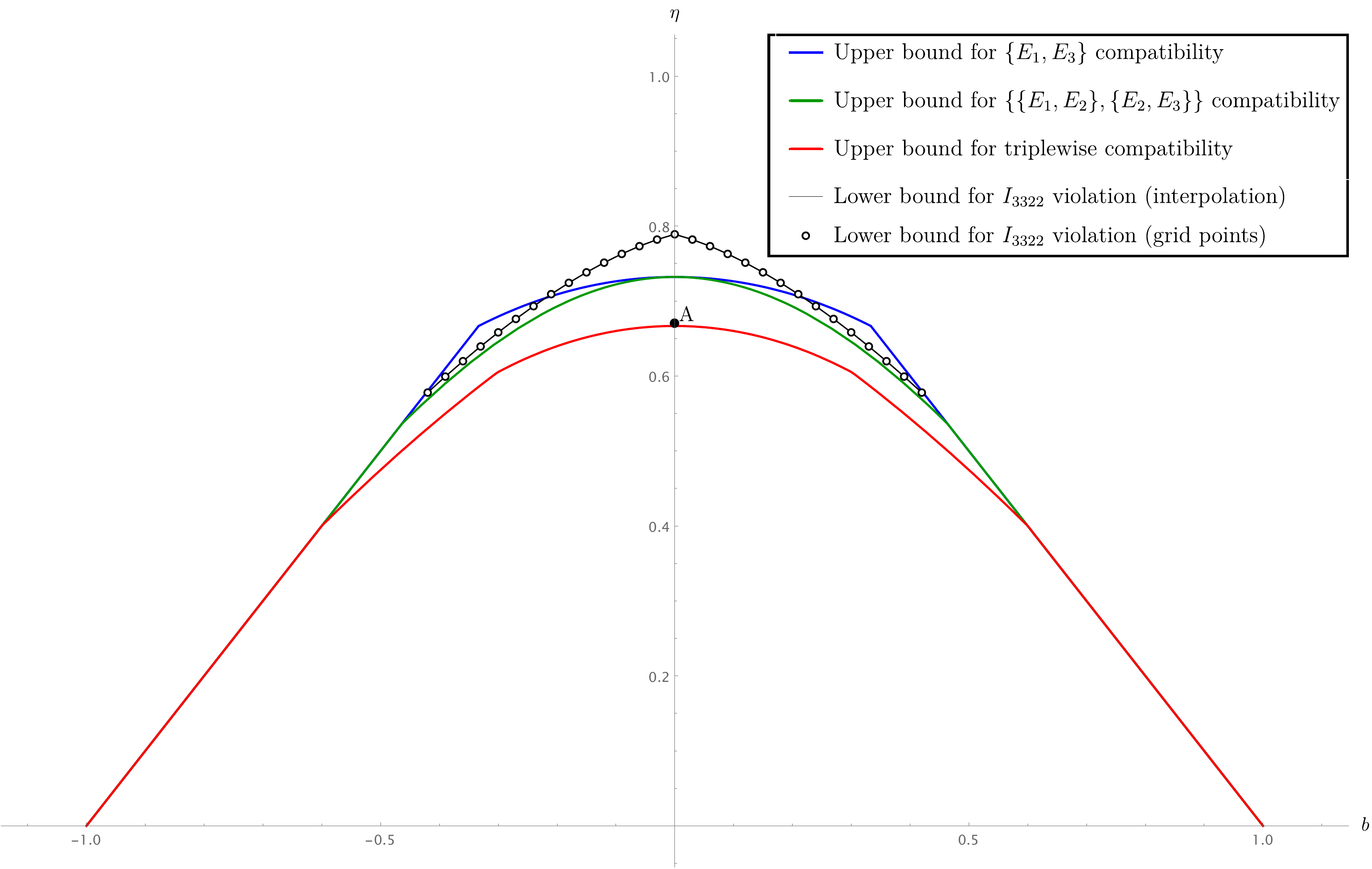}
		\caption{Red line represents the upper bound for $\eta$ given $b$ such that the whole set is compatible. Blue line represents the upper bound for $\eta$ such that $\{E_1,E_3\}$ is compatible. Green line represents the upper bound for $\eta$ such that each of the pairs $\{E_1,E_2\}$ and $\{E_2,E_3\}$ is compatible. Black line represents lower bound for $\eta$ such that we have $I_{3322}$ violation. In the region between the green and the red line, where we have a 3-Specker scenario, there are no $I_{3322}$ violations. However, there are violations in the region between the green and the blue line where only the pair $\{E_1,E_3\}$ is compatible while other two pairs are incompatible. Point $A$ represents the  POVMs used for Alice (same as our Eq.~\eqref{bvexample}) in Ref.~\cite{BV18} to show the that measurement incompatibility does not imply Bell nonlocality.}
		\label{fig:bias2}
	\end{figure*}

	\begin{example}\upshape
		We consider the family in Eq.~\eqref{fam1}, \textit{i.e.},
		\begin{align}
			E_k(x_k)&=\frac{1}{2}\left((1+x_k b)I+\eta x_k\vec{n}_k\cdot\vec{\sigma}\right),\nonumber\\
			\textrm{where }&\vec{n}_k = \cos\frac{2k\pi}{3} \vec{e}_x + \sin\frac{2k\pi}{3} \vec{e}_z, k\in\{1,2,3\}.
		\end{align}
		This set is pairwise jointly measurable if and only if (blue line in Fig.~\ref{fig:bias1}) \cite{YLL10} 
		\begin{equation}
			\eta\leq\begin{cases}
				1+b,\quad -1\leq b <-\frac{1}{3},\\
				\sqrt{3-2b^2}-1,\quad b\in\left[-\frac{1}{3},\frac{1}{3}\right],\\
				1-b,\quad \frac{1}{3}< b\leq1.
			\end{cases}
		\end{equation}
		To determine triplewise joint measurability, we solve an SDP that maximizes $\eta$ for each $b$ picked from a discretization of the interval $[-1,1]$, subject to the existence of a triplewise joint POVM. We obtain numerically a plot for $\eta_1^{(3)}(b)$ such that the set is triplewise jointly measurable iff (red line in Fig.~\ref{fig:bias1})
		\begin{equation}
			\eta\leq\eta_1^{(3)}(b).
		\end{equation}
		Based on our numerical evaluation and visual inspection, we conjecture the exact form of $\eta_1^{(3)}(b)$ to be 
		\begin{equation}
			\eta_1^{(3)}(b)=\begin{cases}
				1+b,\quad -1\leq b<-\frac{1}{3},\\
				\frac{2}{3},\quad b\in\left[-\frac{1}{3},\frac{1}{3}\right],\\
				1-b,\quad \frac{1}{3}< b\leq 1.  
			\end{cases}
		\end{equation}
		This means that the $3$-Specker scenario is realized iff the following system of inequalities is satisfied:
		\begin{equation}
			\abs{b}\leq\frac{1}{3},\quad \frac{2}{3}<\eta\leq\sqrt{3-2b^2}-1.
		\end{equation}
		As is clear in Fig.~\ref{fig:bias1}, there is no violation of the $I_{3322}$ inequality using such measurements when they form a $3$-Specker scenario. 
	\end{example}
	
	\begin{example}\upshape
		We consider the family in Eq.~\eqref{fam2}, \textit{i.e.},
		\begin{align}
			E_k(x_k)&=\frac{1}{2}\left((1+(-1)^k x_k b)I+\eta x_k\vec{n}_k\cdot\vec{\sigma}\right),\nonumber\\
			\textrm{where }&\vec{n}_k = \cos\frac{2k\pi}{3} \vec{e}_x + \sin\frac{2k\pi}{3} \vec{e}_z, k\in\{1,2,3\}.
		\end{align}
		The pair $\{E_1,E_3\}$ is jointly measurable iff (blue line in Fig.~\ref{fig:bias2})
		\begin{equation}
			\eta\leq\begin{cases}
				1+b,\quad -1\leq b<-\frac{1}{3},\\
				\sqrt{3-2b^2}-1,\quad b\in\left[-\frac{1}{3},\frac{1}{3}\right],\\
				1-b,\quad \frac{1}{3}< b\leq 1.
			\end{cases}
		\end{equation}
		while, pairs $\{E_1,E_2\}$ and $\{E_2,E_3\}$ are jointly measurable iff (green line in Fig.~\ref{fig:bias2})
		\begin{equation}
			\eta\leq\begin{cases}
				1+b,\quad -1\leq b<3-2\sqrt{3},\\
				\sqrt{2}\sqrt{2+b^2-\sqrt{3(1+2b^2)}},\quad \abs{b}\leq 2\sqrt{3}-3\\
				1-b,\quad 2\sqrt{3}-3<b\leq1.
			\end{cases}
		\end{equation}
		We solve an SDP---that maximizes $\eta$ for each $b$ picked from a discretization of the interval $[-1,1]$, subject to the existence of a triplewise joint POVM---to obtain $\eta_2^{(3)}(b)$ such that we have triplewise joint measurability iff (red line in Fig.~\ref{fig:bias2})
		\begin{equation}
			\eta\leq\eta_2^{(3)}(b).
		\end{equation}
		As is clear in Fig.~\ref{fig:bias2}, there is no violation of the $I_{3322}$ inequality using such measurements when they form a $3$-Specker scenario.	
	\end{example}
	
\begin{figure*}[!htb]
		\centering
		\includegraphics[scale=0.3]{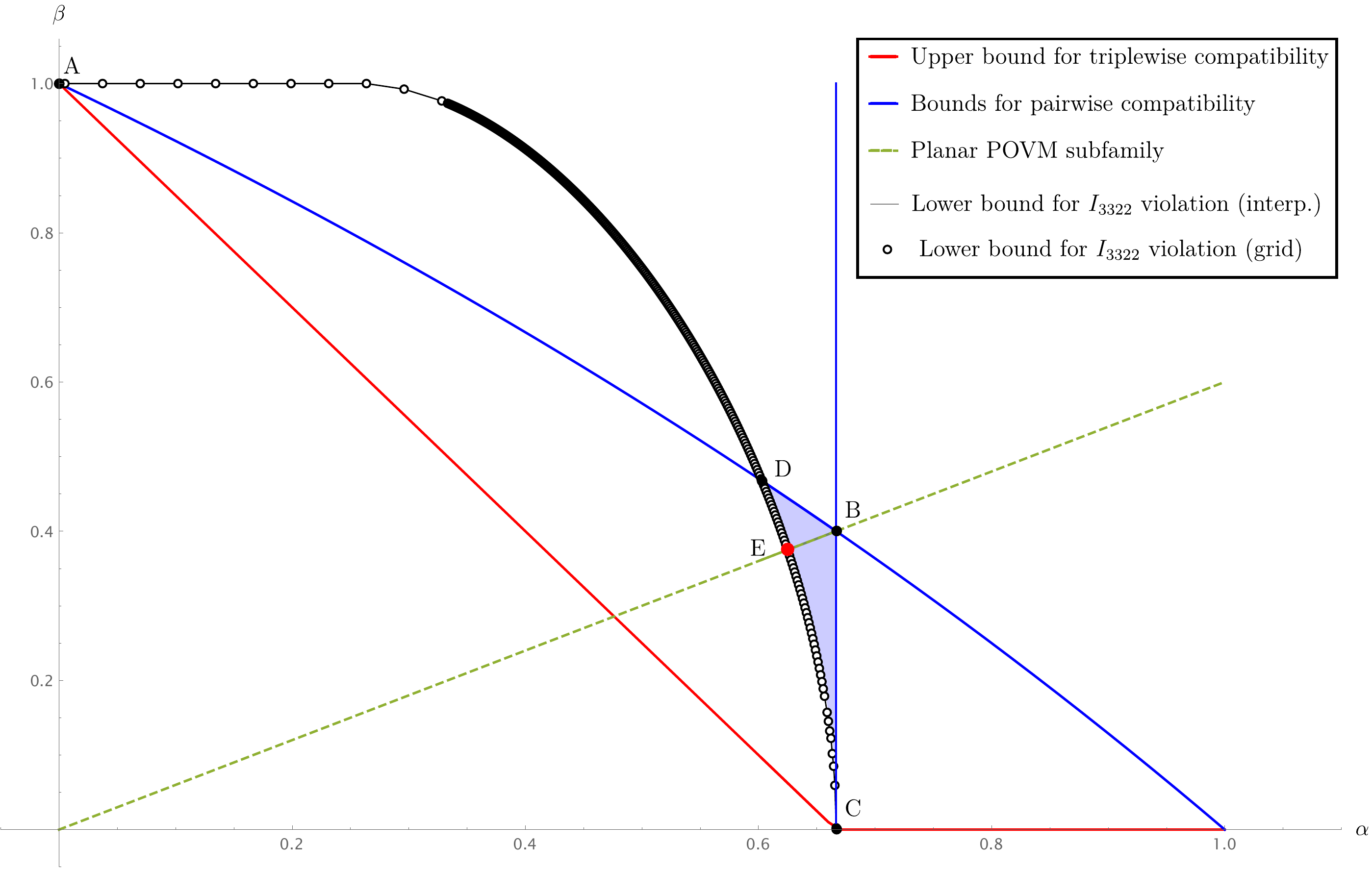}
		\caption{Red line represents upper bound for $\beta$ such that we have 3-way compatibility. Blue line passing through points A and B represents the upper bound for $\beta$ such that $\{E_1,E_3\}$ and $\{E_2,E_3\}$ are compatible. Vertical blue line represents the upper bound for alpha such that the pair $\{E_1,E_2\}$ is compatible. For triangle-like region ABC, 3-Specker is realized. Black line represents the lower bound for $\beta$ given $\alpha$ such that we have $I_{3322}$ violation. In the region defined by the points BCD we have a 3-Specker scenario capable of violating $I_{3322}$ Bell inequality. Dashed line represents the planar family of POVMs from Appendix B in Ref.~\cite{QVB14}. The red dot on that line is the critical point for $I_{3322}$ violation by the family in the same Appendix (its criticality is here confirmed).}
		\label{fig:famquint}
	\end{figure*}

	\begin{example}\upshape
		We consider the family given in Eq.~\eqref{quintino}. Pairs $\{E_1,E_3\}$ and $\{E_2,E_3\}$ are jointly measurable iff (blue diagonal line in Fig.~\ref{fig:famquint})
		\begin{equation}
			\beta\leq\frac{4-4\alpha}{4-\alpha}.
		\end{equation}
		The pair $\{E_1,E_2\}$ is jointly measurable iff (blue vertical line in Fig~\ref{fig:famquint})
		\begin{equation}
			\alpha\leq\frac{2}{3}.
		\end{equation}
		From an SDP---maximizing $\beta$ for each value of $\alpha$ picked from a discretization of the interval $[0,1]$ subject to the existence of a triplewise joint POVM---we obtain the boundary for triplewise joint measurability, \textit{i.e.}, our family is compatible iff
		\begin{equation}
			\beta\leq\beta^{(3)}(\alpha).
		\end{equation}
		Based on our numerical evaluation and visual inspection, we conjecture the boundary $\beta^{(3)}(\alpha)$ (red line in Fig~\ref{fig:famquint}) to be
		\begin{equation}
			\beta^{(3)}(\alpha) = \begin{cases}
				1-\frac{3}{2}\alpha,\textrm{ for } \alpha<\frac{2}{3},\\
				0,\textrm{ for }\frac{2}{3}\leq\alpha\leq1
			\end{cases}.
		\end{equation}
		We refer to Fig.~\ref{fig:famquint} for a depiction of the region (``BCD") where the $I_{3322}$ inequality is violated by a $3$-Specker scenario.
	\end{example}
	
\bibliographystyle{unsrturl}

\end{document}